\documentclass[aip,jcp,reprint,floatfix]{revtex4-1}
\usepackage{amsmath,amssymb}
\usepackage{epsfig}
\usepackage{graphicx}
\usepackage{dcolumn}
\usepackage{bbold}
\usepackage{natbib}
\usepackage[normalem]{ulem} 
\usepackage[colorlinks=true,citecolor={blue},urlcolor={blue}, linkcolor=blue]{hyperref}%

\newcommand{\ue}[1]{$^{\textrm{#1}}$}

\newcommand{\dtp}{\partial^{\hspace{.03em}\prime}_t}
\newcommand{\varp}{\delta^{\hspace{.03em}\prime}}

\begin{document}


\title{Time-dependent optimized coupled-cluster method for multielectron dynamics
}



\author{Takeshi Sato}
\email[Electronic mail:]{sato@atto.t.u-tokyo.ac.jp}
\affiliation{
Photon Science Center, School of Engineering, 
The University of Tokyo, 7-3-1 Hongo, Bunkyo-ku, Tokyo 113-8656, Japan
}
\affiliation{
Department of Nuclear Engineering and Management, School of Engineering,
The University of Tokyo, 7-3-1 Hongo, Bunkyo-ku, Tokyo 113-8656, Japan
}
\author{Himadri Pathak}
\email[Electronic mail:]{pathak@atto.t.u-tokyo.ac.jp}
\affiliation{
Department of Nuclear Engineering and Management, School of Engineering,
The University of Tokyo, 7-3-1 Hongo, Bunkyo-ku, Tokyo 113-8656, Japan
}
\author{Yuki Orimo}
\email[Electronic mail:]{ykormhk@atto.t.u-tokyo.ac.jp}
\affiliation{
Department of Nuclear Engineering and Management, School of Engineering,
The University of Tokyo, 7-3-1 Hongo, Bunkyo-ku, Tokyo 113-8656, Japan
}
\author{Kenichi L. Ishikawa}
\email[Electronic mail:]{ishiken@n.t.u-tokyo.ac.jp}
\affiliation{
Department of Nuclear Engineering and Management, School of Engineering,
The University of Tokyo, 7-3-1 Hongo, Bunkyo-ku, Tokyo 113-8656, Japan
}
\affiliation{
Photon Science Center, School of Engineering, 
The University of Tokyo, 7-3-1 Hongo, Bunkyo-ku, Tokyo 113-8656, Japan
}



\begin{abstract}
Time-dependent coupled-cluster method with time-varying
orbital functions, called time-dependent optimized coupled-cluster (TD-OCC) method,
is formulated for multielectron dynamics in an
intense laser field.
We have successfully derived equations of motion for CC amplitudes and orthonormal orbital functions
based on the real action functional, and implemented the method including double excitations (TD-OCCD) and double and triple excitations (TD-OCCDT)
within the optimized active orbitals. 
The present method is size extensive and gauge invariant, a polynomial cost-scaling alternative to the 
time-dependent multiconfiguration self-consistent-field method. 
The first application of the TD-OCC method to intense-laser driven correlated electron dynamics in Ar atom is reported.
\end{abstract}


\maketitle 


\section*{introduction}\label{sec:introduction}
The strong-field physics and ultrafast science are rapidly progressing
towards a world-changing goal to directly measure and control the electron motion in
matters.\cite{Protopapas1997RPP,Agostini2004RPP,Krausz2009RMP,Gallmann2013ARPC,Nisoli:2017} Reliable theoretical and computational
methods are indispensable to understand and predict strong-field and ultrafast phenomena,
which frequently involve both bound excitations and ionizations,
where dynamical electron correlation plays a key role.

One of the most advanced methods to describe such electron dynamics is the
multiconfiguration time-dependent Hartree-Fock (MCTDHF) method,\cite{Zanghellini:2003,Kato:2004,Caillat:2005,Nest:2005a} 
and more generally the time-dependent multiconfiguration self-consistent-field (TD-MCSCF) method.\cite{Nguyen-Dang:2007,Ishikawa:2015} 
In TD-MCSCF, the electronic wavefunction is given by the
configuration-interaction (CI) expansion, $\Psi_{\rm CI}(t)=\sum_{\bf I}C_{\bf I}(t)\Phi_{\bf I}(t)$, 
and both CI coefficiets $\{C_{\bf I}(t)\}$ and {\it occupied} spin-orbital functions $\{\psi_p(t)\}$
constituting the Slater determinants $\{\Phi_{\bf I}(t)\}$ are propagated in time.
(Hereafter spin-orbital functions are simply referred to as orbitals.
See, e.g,
Refs.~\citenum{Rohringer2006PRA,Pabst:2013,Hochstuhl:2012,Bauch:2014}
for TDCI methods using fixed orbitals, and Ref.~\citenum{Ishikawa:2015}
for a broad review of {\it ab initio} wavefunction-based methods for
multielectron dynamics.)
Though powerful, the full CI-based MCTDHF method suffers from the factorial scaling of the computational cost
with respect to the number of electrons,
which restricts its applicability to the systems of moderate size.

\begin{figure}[!b]
\centering
\includegraphics[width=0.9\linewidth,clip]{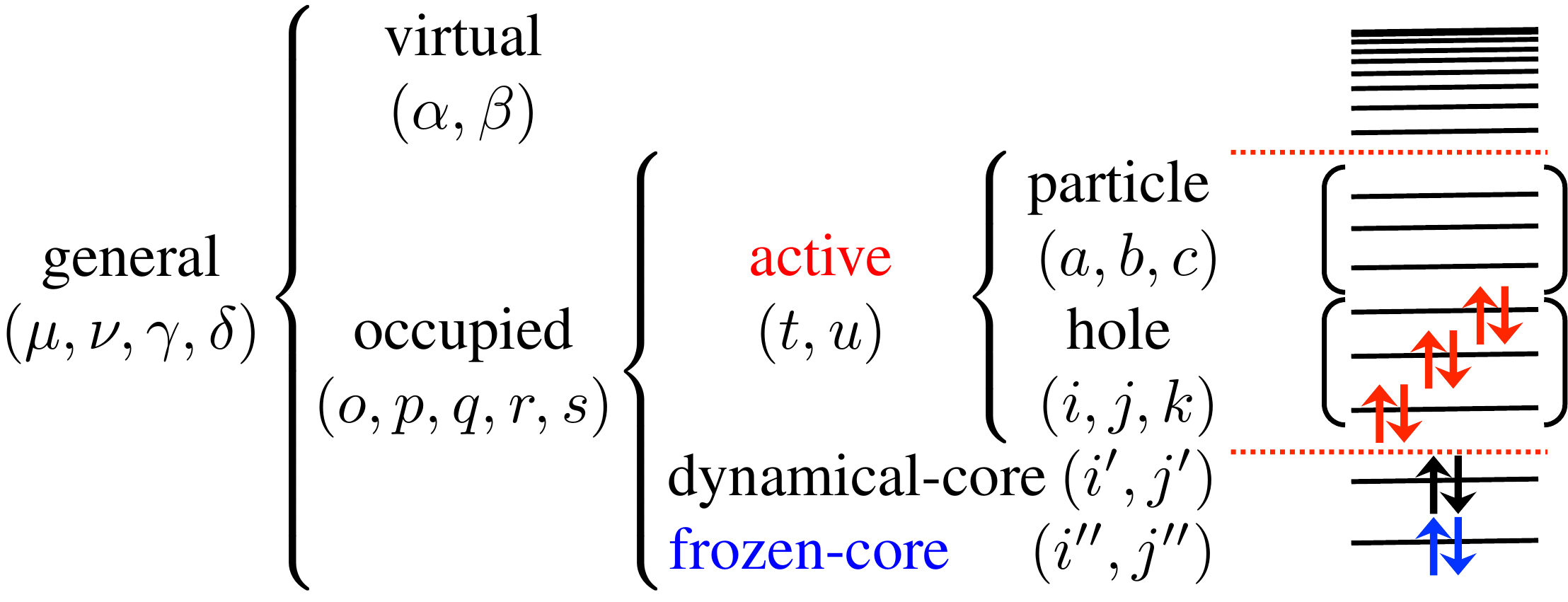}
\caption{\label{fig:orbitalspace}
Illustration of the orbital subspacing (for a spin-degenerate case). 
The up and down arrows represent
electrons, and 
horizontal lines represent
spatial orbitals, classified into frozen-core (doubly occupied and fixed in time) and dynamical-core
(doubly occupied but propagated in time) orbitals and active
orbitals (correlated and propagated). The active orbital space is further splitted into the hole 
and particle subspaces. 
The index convention for orbitals of each group is given in the parentheses. 
}%
\end{figure}

The time-dependent complete-active-space self-consistent-field (TD-CASSCF) method\cite{Sato:2013} significantly reduces the cost by 
flexibly classifying occupied orbitals into {\it frozen-core},
{\it dynamical-core}, 
and {\it active} orbitals as defined in Fig.~\ref{fig:orbitalspace}. 
More approximate and thus less demanding models have been actively investigated,
\cite{Miyagi:2013,Miyagi:2014b,Haxton:2015,Sato:2015} which rely on the truncaed CI expansion
within the active orbitals given, e.g, as
\begin{eqnarray}\label{eq:ciwfn}
\Psi_{\rm CI}
= (C_0+C^a_i\hat{E}^a_i+C^{ab}_{ij}\hat{E}^{ab}_{ij}+C^{abc}_{ijk}\hat{E}^{abc}_{ijk}+\cdots)\Phi,
\end{eqnarray}
where (Einstein's summation convention is implied throughout for orbital indices.) 
$\hat{E}^{abc\cdots}_{ijk\cdots}=\hat{c}^\dagger_{a}\hat{c}^\dagger_{b}\hat{c}^\dagger_{c}\cdots\hat{c}_{k}\hat{c}_{j}\hat{c}_{i}$
[$c^\dagger_\mu$ ($c_\mu$) is a creation (anihilation) operator for
$\psi_\mu$] excites electron(s) from orbital(s) $i,j,k\cdots$ occupied in the
reference determinant $\Phi$ (hole) to those $a,b,c\cdots$ not occupied
in the reference (particle), achieving a polynomial cost scaling.
Even though state-of-the-art real-space implementations\cite{Kato:2008,Hochstuhl:2011,Haxton:2012,Haxton:2014,Haxton:2015,Sato:2016,Sawada:2016,Omiste:2017,Orimo:2018}
have proved a great utility of these methods,
all the methods based on a truncated-CI expansion share a common drawback of
not being size extensive.\cite{Szabo:1996,Helgaker:2002}

Therefore one naturally seeks an alternative to Eq.~(\ref{eq:ciwfn}), where the truncated CI is
replaced with the size-extensize coupled-cluster (CC) 
expansion\cite{Szabo:1996,Helgaker:2002,Kummel:2003,Shavitt:2009}
\begin{eqnarray}\label{eq:ccwfn}
\Psi_{\rm CC} = \exp(\tau^a_i\hat{E}^a_i+\tau^{ab}_{ij}\hat{E}^{ab}_{ij}+\tau^{abc}_{ijk}\hat{E}^{abc}_{ijk}+\cdots)\Phi,
\end{eqnarray}
where both excitation amplitudes $\{\tau^{abc\cdots}_{ijk\cdots}(t)\}$ and orbitals are propagated in time, 
just as the stationary {\it orbital-optimized} CC (OCC) method\cite{Scuseria:1987,Sherrill:1998,Krylov:1998,Kohn:2005}
optimizes both excitation amplitudes and orbitals to minimize the CC energy functional.
This idea was recently put forward by Kvaal\cite{Kvaal:2012}, who, based on Arponen's seminal work,\cite{Arponen:1983} developed a time-dependent coupled-cluster
method using time-varying biorthonormal orbitals, called orbital-adapted time-dependent coupled-cluster
(OATDCC) method, and applied it to the collision problem in a one-dimensional model Hamiltonian. 
This method, however, has a difficulty in obtaining the ground state of the system, required as an initial state of simulations, 
and has not yet been applied to real three-dimensional or time-dependent Hamiltonian.
We also notice recent investigations on the time-dependent coupled-cluster methods using fixed orbitals.\cite{Huber:2011,Nascimento:2016}
Their primary focus, however, has been placed on retrieving spectral information rather than the time-domain electron dynamics itself.

In this communication we report our successful formulation and implementation of
the time-dependent coupled-cluster method using time-varying orthonormal orbitals,
called time-dependent optimized coupled-cluster (TD-OCC) method, 
and present its first application to electron dynamics of Ar atom subject to an intense laser field.
The Hartree atomic units are used throughout unless otherwise noted.

\section*{Time-dependent variational principle with real CC action functional\label{sec:general}}%
We are interested in the dynamics of a many-electron system
interacting with external electromagnetic fields governed by the Hamiltonian written in the second quantization as
\begin{eqnarray}\label{eq:ham}
&&\quad\quad\quad\quad  
\hat{H}(t)=h^\mu_\nu(t)\hat{E}^\mu_\nu+\frac{1}{2}g^{\mu\gamma}_{\nu\delta}\hat{E}^{\mu\gamma}_{\nu\delta},
 \\
&&\quad\quad\quad h^\mu_\nu(t)=\int dx\psi^*_\mu [h_0+V_{\rm ext}(t)]\psi_\nu, \\
&&g^{\mu\gamma}_{\nu\delta}=\int\int dx_1 dx_2\psi^*_\mu(x_1)\psi^*_\gamma(x_2) r^{-1}_{12}\psi_\nu(x_1)\psi_\delta(x_2),
\end{eqnarray}
where 
[$x\!=\!(\pmb{r},\sigma)$ labels
spatial-spin coordinate.] $h_0$ and $V_{\rm ext}$ are the field-free one-electron Hamiltonian and
the interaction of an electron and external fields, given, e.g, in the electric dipole approximation as
$V_{\rm ext}=\pmb{E}(t)\cdot\pmb{r}$ in the length gauge or
$V_{\rm ext}=-i\pmb{A}(t)\cdot\pmb{\nabla}_{\pmb{r}}$ in the
velocity gauge, with $\pmb{E}(t)$ and $\pmb{A}(t)=-\int^t dt^\prime
\pmb{E}(t^\prime)$ being the electric field and vector potential,
respectively.

Note that we include in Eq.~(\ref{eq:ham}) not only a given number $N_o$ of occupied orbitals
$\{\psi_p\}$ 
but also $N_v$ {\it virtual} orbitals $\{\psi_\alpha\}$ 
which are orthogonal to $\{\psi_p\}$. The number $N=N_o+N_v$ of all
orbitals $\{\psi_\mu\}$ is equal to the number of basis functions to
expand orbitals (Fig~\ref{fig:orbitalspace}), and in the basis-set limit ($N\rightarrow\infty$), $\hat{H}$ of Eq.~(\ref{eq:ham}) is equivalent to the first-quantized
Hamiltonian. The index convention given in Fig.~\ref{fig:orbitalspace} is used for orbitals.

Following the previous work\cite{Pedersen:1999,Kvaal:2012,Helgaker:2002}, we begin with the
coupled-cluster Lagrangian $L$,
\begin{eqnarray}\label{eq:lagrangian}
L &=& \langle\Psi_{\rm L}|(\hat{H}-i{\partial_t})|\Psi_{\rm R}\rangle \nonumber \\
&=& \langle\Phi|(1+\hat{\Lambda})e^{-\hat{T}}
(\hat{H}-i{\partial_t})
e^{\hat{T}}|\Phi\rangle, \\
\hat{T} &=& \label{eq:tamp}
\tau_i^a \hat{E}_i^a +
\tau_{ij}^{ab} \hat{E}_{ij}^{ab} +
\tau_{ijk}^{abc} \hat{E}_{ijk}^{abc} + \cdots = \tau_I^A \hat{E}_I^A \nonumber \\
&=& \hat{T}_1+\hat{T}_2+\hat{T}_3+\cdots, \\
\hat{\Lambda} &=&  \label{eq:lamp}
\lambda^i_a \hat{E}^i_a +
\lambda^{ij}_{ab} \hat{E}^{ij}_{ab} +
\lambda^{ijk}_{abc}\hat{E}^{ijk}_{abc} + \cdots  = \lambda^I_A \hat{E}^I_A \nonumber \\
&=& \hat{\Lambda}_1+\hat{\Lambda}_2+\hat{\Lambda}_3+\cdots,
\end{eqnarray}
where
$|\Phi\rangle=\prod_{i^{\prime\prime}i^{\prime}i}\hat{c}^\dagger_{i^{\prime\prime}}\hat{c}^\dagger_{i^\prime}\hat{c}^\dagger_{i}|\rangle$ 
is a reference determinant.
For the notational brevity, we use composite indices $I=ijk\cdots$ and $A=abc\cdots$ to denote general-rank excitation
(deexcitation) operators and amplitudes as $\hat{E}^A_I$ ($\hat{E}^I_A$)
and $\tau^A_I$ ($\lambda^I_A$). Note that we use the orthonormal orbitals $\{\psi_\mu\}$ and
$\langle\Phi|=|\Phi\rangle^\dagger$.

As the guiding principle,
we adopt the manifestly real action functional, 
\begin{eqnarray}
S &=& \label{eq:action}
\Re\int_{t_0}^{t_1} Ldt = \frac{1}{2} \int_{t_0}^{t_1}\left( L + L^*\right) dt,
\end{eqnarray}
and require the action to be stationary
\begin{eqnarray}
\delta S &=& \label{eq:vars0}
\langle\delta\Psi_{\rm L}|\hat{H}|\Psi_{\rm R}\rangle
+\langle\Psi_{\rm L}|\hat{H}|\delta\Psi_{\rm R}\rangle \nonumber \\
&-&i\langle\delta\Psi_{\rm L}|\dot{\Psi}_{\rm R}\rangle
+i\langle\dot{\Psi}_{\rm L}|\delta\Psi_{\rm R}\rangle + c.c = 0,
\end{eqnarray}
with respect to the variation of left- and right-state wavefunctins.
This real-action formulation using orthonormal orbitals 
(also adopted in Ref.~\citenum{Pedersen:1999} for a gauge-invariant response theory)
distinguishes our method from the
OATDCC method,\cite{Kvaal:2012} which is based on the bivariational principle with the complex-analytic action functional $S=\int dt L$ using biorthonormal orbitals.

The time derivative and variation of the left- and right-state wavefunctions can be written as\cite{Miranda:2011a,Sato:2013,Miyagi:2014b,Sato:2015}
\begin{subequations} \label{eq:dtandvar}
\begin{eqnarray}
&|\dot{\Psi}_{\rm R}\rangle = \label{eq:dtket}
(\dtp\hat{T} +  \hat{X})e^{\hat{T}}|\Phi\rangle,& \\
&|\delta\Psi_{\rm R}\rangle = \label{eq:varket}
(\varp\hat{T} + \hat{\Delta})e^{\hat{T}}|\Phi\rangle,& \\
&\langle\dot{\Psi}_{\rm L}| =\label{eq:dtbra}
 \langle\Phi|[-(1+\Lambda)e^{-\hat{T}}(\dtp\hat{T}+\hat{X})+\dtp\hat{\Lambda}e^{-\hat{T}}],& \\
&\langle\delta\Psi_{\rm L}| =\label{eq:varbra}
 \langle\Phi|[-(1+\Lambda)e^{-\hat{T}}(\varp\hat{T}+\hat{\Delta})+\varp\hat{\Lambda}e^{-\hat{T}}],&
\end{eqnarray}
\end{subequations}
where 
$\dtp\hat{T}=\dot{\tau}^A_I\hat{E}^A_I$,
$\varp\hat{T}=\delta\tau^A_I\hat{E}^A_I$,
$\dtp\hat{\Lambda}=\dot{\lambda}^I_A\hat{E}^I_A$,
$\varp\hat{\Lambda}=\delta\lambda^I_A\hat{E}^I_A$,
$\hat{X}=X^\mu_\nu\hat{E}^\mu_\nu$,
$\hat{\Delta}=\Delta^\mu_\nu\hat{E}^\mu_\nu$, 
$X^\mu_\nu = \langle\psi_\mu|\dot{\psi}_\nu\rangle$, and
$\Delta^\mu_\nu = \langle\psi_\mu|\delta{\psi}_\nu\rangle$.
The matrices $X$ and $\Delta$
are anti-Hermitian to enforce the orthonormality of orbitals
$\langle\psi_\mu(t)|\psi_\nu(t)\rangle=\delta^\mu_\nu$ at any time.
We insert Eqs.~(\ref{eq:dtandvar}) into Eq.~(\ref{eq:vars0}) and
group terms for each kind of variation to obtain

\begin{widetext}
\begin{subequations} \label{eq:vars}
\begin{eqnarray}
\delta S 
&=&\label{eq:vars_amp}
\delta\tau^A_I\left\{
\langle\Phi|(1+\Lambda)e^{-\hat{T}}
[\hat{H}-i\hat{X},\hat{E}^A_I]e^{\hat{T}}|\Phi\rangle
+i\dot{\lambda}^I_A\right\} 
+\delta\lambda^I_A\left\{
\langle\Phi^A_I|e^{-\hat{T}}
\left(\hat{H}-i\hat{X}\right)e^{\hat{T}}|\Phi\rangle
-i\dot{\tau}^A_I\right\} + c.c \\
&+&
\Delta^{\mu}_{\nu} \left\{
\langle\Phi|(1+\Lambda)e^{-\hat{T}}[\hat{H}-i(\dtp\hat{T})-i\hat{X},\hat{E}^{\mu}_{\nu}]e^{\hat{T}}|\Phi\rangle
+i\langle\Phi|(\dtp\hat{\Lambda})e^{-\hat{T}}\hat{E}^{\mu}_{\nu}e^{\hat{T}}|\Phi\rangle
\right. \nonumber \\
&& \label{eq:vars_orb}
\hspace{1.3em} \left.-
\langle\Phi|(1+\Lambda)e^{-\hat{T}}[\hat{H}-i(\dtp\hat{T})-i\hat{X},\hat{E}^{\nu}_{\mu}]e^{\hat{T}}|\Phi\rangle^*
-i\langle\Phi|(\dtp\hat{\Lambda})e^{-\hat{T}}\hat{E}^{\nu}_{\mu}e^{\hat{T}}|\Phi\rangle^*
\right\},
\end{eqnarray}
\end{subequations}
\end{widetext}
where the anti-Hermiticity of $\Delta$ is
used for deriving Eq.~(\ref{eq:vars_orb}). 
The EOMs for the amplitudes
$\tau^A_I$ and $\lambda^I_A$ are readily derived from the
conditions $\delta S/\delta \lambda^I_A=0$ and $\delta S/\delta \tau^A_I=0$, 
respectively, as
\begin{eqnarray}
i\dot{\tau}^A_I &=& \label{eq:eom_t}
\langle\Phi_I^A| e^{-\hat{T}} (\hat{H}-i\hat{X})e^{\hat{T}}|\Phi\rangle, \\
-i\dot{\lambda}^I_A &=& \label{eq:eom_l}
\langle\Phi|(1+\hat{\Lambda}) e^{-\hat{T}} [\hat{H}-i\hat{X},\hat{E}^A_I] e^{\hat{T}}|\Phi\rangle.
\end{eqnarray}

The EOMs for orbitals are to be obtained from $\delta S/\delta \Delta^\mu_\nu=0$.
After straightforward yet tedious algebraic
manipulations\cite{Sato:2018}, we found that (i) orbital pairs within the same
orbital space (with no further subclassification in Fig.~\ref{fig:orbitalspace})
are redundant as well known for the stationary single-reference CC methods, i.e,
$X^{i^{\prime\prime}}_{j^{\prime\prime}},X^{i^{\prime}}_{j^{\prime}},X^i_j,X^a_b$ can be arbitrary anti-Hermition
matrix elements, and (ii) the expressions of all the other $X^\mu_\nu$ terms except
the hole-particle contributions $X^a_i$ (and $X^i_a=-X^{a*}_i$) are formally identical to those for the TD-CASSCF method.\cite{Sato:2013}
%
%
Thus we follow the discussion in Ref.~\citenum{Sato:2013}, and write
the final expression for the orbital EOM, with the hole-particle
terms $\{X^a_i\}$ left unspecified until next section, 
\begin{eqnarray}
&i|\dot{\psi_p}\rangle = \label{eq:eom_orb}
(1-\hat{P}) \hat{F}_p|\psi_p\rangle + i|\psi_q\rangle X^q_p,& \\
&\hat{F}_p|\psi_p\rangle = \label{eq:eom_gfockoperator}
\hat{h} |\psi_p\rangle + \hat{W}^r_s|\psi_q\rangle P^{qs}_{or} (D^{-1})^o_p,& \\
&D^p_q=\label{eq:rdmh}\frac{1}{2}(\rho^p_q+\rho^{q*}_p), \hspace{.5em}
P^{pr}_{qs}=\frac{1}{2}(\rho^{pr}_{qs}+\rho^{qs*}_{pr}),& \\
&\rho^p_q = \label{eq:1rdm}
\langle\Phi|(1+\Lambda)e^{-\hat{T}}\hat{E}^q_pe^{\hat{T}}|\Phi\rangle,& \\
&\rho^{pr}_{qs} = \label{eq:2rdm}
\langle\Phi|(1+\Lambda)e^{-\hat{T}}\hat{E}^{qs}_{pr}e^{\hat{T}}|\Phi\rangle,&
\end{eqnarray}
where $\hat{P}=\sum_p|\psi_p\rangle\langle\psi_p|$, 
$\hat{W}^r_s$ is the mean-field operator\cite{Sato:2013},
and $\{X^p_q\}$ except $\{X^a_i\}$ are given in Ref.~\citenum{Sato:2013}. 
(Note that the Hermitian matrix $R\equiv iX$ was
used as the working variable in Ref.\citenum{Sato:2013}.
See also Ref.~\citenum{Sato:2016} for the velocity gauge treatment of frozen-core
orbitals.) We note the natural appearance of the Hermitialized reduced density matrices (RDMs) $D$
and $P$ is the direct consequece of relying on the real action functional with orthonormal orbitals.

\begin{table}[!b]
\caption{\label{tab:enemol}
Total energies (Hartree) of BH molecule (the bond length 2.4 \AA) with
the DZP basis set.
}
\begin{ruledtabular}
\begin{tabular}{lcc}
&
\multicolumn{1}{c}{This work\footnotemark[1]} &
\multicolumn{1}{c}{Ref.~\citenum{Krylov:1998}} \\
\hline
HF     & -25.124\hspace{.25em}742\hspace{.25em}010 & -25.124\hspace{.25em}742 \\
OCCD   & -25.178\hspace{.25em}285\hspace{.25em}704 & -25.178\hspace{.25em}286 \\
OCCDT  & -25.178\hspace{.25em}312\hspace{.25em}565 &  \\
CASSCF & -25.178\hspace{.25em}334\hspace{.25em}889 & -25.178\hspace{.25em}335 \\
\end{tabular}
\end{ruledtabular}
\footnotetext[1]{The overlap, one-electron, and two-electron repulsion integrals over gaussian basis functions are generated using
Gaussian09 program (Ref.~\citenum{Frisch:2009}), and used to propagate EOMs in imaginary time in the orthonormalized gaussian basis, 
with a convergence threshold of 10\ue{-15} Hartree of energy difference in subsequent time steps.}
\end{table}

\section*{TD-OCC method\label{sec:tdocc}}
In order to derive a working equation for the hole-particle rotations
$\{X^a_i\}$, and thus complete the derivation of EOMs, one has to specify a
particular ansatz of the CC Lagrangian, i.e., which terms to be included
in Eqs.~(\ref{eq:tamp}) and (\ref{eq:lamp}). We, therefore, define the TD-OCC method by the ansatz
\begin{eqnarray}\label{eq:ansatz_occ}
\hat{T} = \hat{T}_2+\hat{T}_3+\cdots, \hspace{.5em}
\hat{\Lambda} = \hat{\Lambda}_2+\hat{\Lambda}_3+\cdots, 
\end{eqnarray}
that is, with single amplitudes $\hat{T}_1$ and $\hat{\Lambda}_1$ omitted.
The same form has been adopted in the ground-state OCC methods\cite{Scuseria:1987,Sherrill:1998,Krylov:1998} and OATDCC method. 
We have also tried to retain single amplitudes ($\hat{T}=\hat{T}_1+\hat{T}_2+\cdots$, 
$\hat{\Lambda}=\hat{\Lambda}_1+\hat{\Lambda}_2+\cdots$) and apply TDVP of the previous section. 
However, the resultant EOMs were stable neither in real time nor imaginary time propagation.\cite{Sato:2018}
A similar difficulty has been reported for the stationary OCC
method,\cite{Scuseria:1987} which is in essence due to the similar physical
roles played by single excitations and hole-particle rotations. 

For the ansatz of Eq.~(\ref{eq:ansatz_occ}), we derive the system of equations to be solved for $X^a_i$
[by coupling Eqs.~(\ref{eq:eom_t}), (\ref{eq:eom_l}) and
$\delta S/\delta \Delta^i_a = 0$],\cite{Sato:2018} as
\begin{eqnarray}
&i\left\{(\delta^a_bD^j_i-D^a_b\delta^j_i)X^b_j+\frac{1}{2}A^{aj}_{ib}  X^{b*}_j\right\} = \label{eq:occ_eom_hp}
B^a_i,& \\
&A^{aj}_{ib} = \label{eq:amat}
\langle\Phi|(1+\hat{\Lambda}) e^{-\hat{T}} [\hat{E}^A_I,\hat{E}^j_b] e^{\hat{T}}|\Phi\rangle
\langle\Phi^A_I|e^{-\hat{T}}\hat{E}^{i}_{a}e^{\hat{T}}|\Phi\rangle& 
\nonumber \\ &-
\langle\Phi|(1+\Lambda)e^{-\hat{T}}[\hat{E}^A_I,\hat{E}^{i}_{a}]e^{\hat{T}}|\Phi\rangle
\langle\Phi_I^A| e^{-\hat{T}} \hat{E}^j_be^{\hat{T}}|\Phi\rangle,& \\
&B^a_i = \label{eq:bvec}
F^a_jD^j_i - D^a_bF^{i*}_b& \\
&\quad+
\frac{1}{2}\langle\Phi|(1+\hat{\Lambda}) e^{-\hat{T}} [\hat{E}^A_I,\hat{H}] e^{\hat{T}}|\Phi\rangle
\langle\Phi^A_I|e^{-\hat{T}}\hat{E}^{i}_{a}e^{\hat{T}}|\Phi\rangle&
\nonumber \\ 
&\quad-
\frac{1}{2}\langle\Phi|(1+\Lambda)e^{-\hat{T}}[\hat{E}^A_I,\hat{E}^{i}_{a}]e^{\hat{T}}|\Phi\rangle
\langle\Phi_I^A| e^{-\hat{T}} \hat{H}e^{\hat{T}}|\Phi\rangle.& \nonumber
\end{eqnarray}
A special simplification arises when only double amplitudes are
included, in which case Eq.~(\ref{eq:amat}) and the last two terms of
Eq.~(\ref{eq:bvec}) vanish, and $\hat{X}$ needs not be included in
Eqs.~(\ref{eq:eom_t}) and (\ref{eq:eom_l}). 

In summary, the TD-OCC EOMs consist of
Eq.~(\ref{eq:eom_t}) for $\hat{T}_2,\hat{T}_3,\cdots$, 
Eq.~(\ref{eq:eom_l}) for $\hat{\Lambda}_2,\hat{\Lambda}_3\cdots$, and
Eq.~(\ref{eq:eom_orb}) for orbitals, with $\{X^a_i\}$ obtained by
solving Eq.~(\ref{eq:occ_eom_hp}). 
The present TD-OCC method is, as the name suggests, a direct time-dependent extension of
the stationary OCC method\cite{Scuseria:1987,Sherrill:1998,Krylov:1998}.

\section*{Feasibility of imaginary relaxation method\label{sec:imag}}
We have implemented the TD-OCC method with double excitations
(TD-OCCD) and double and triple excitations (TD-OCCDT) for atom-centered gaussian basis functions
and the 
atomic Hamiltonian with finite-element
discrete variable representation (FEDVR) basis\cite{Sato:2016,Orimo:2018}. These
codes share the same implementation for the basis-independent
procedures [Eqs.~(\ref{eq:eom_t}), (\ref{eq:eom_l}),
(\ref{eq:rdmh})-(\ref{eq:2rdm}), (\ref{eq:occ_eom_hp})-(\ref{eq:bvec})].

It turns out, in contrast to the OATDCC method, that the
imaginary-time relaxation is feasible for the TD-OCC method to
obtain the ground state, which is quite convenient when using 
grids or locally-supported basis like FEDVR to expand orbitals.
This is confirmed in table~\ref{tab:enemol}, which compares the
total energies of the ground state of BH molecule with
double-$\zeta$ plus polarization (DZP) basis\cite{Harrison:1983} obtained by imaginary
time propagation 
with results of corresponding stationary methods\cite{Krylov:1998}.
The OCC and CASSCF methods correlate all six electrons among six active
orbitals.
The agreement of OCCD energies up to the reported (six) decimal places in
Ref.~\citenum{Krylov:1998} clearly demonstrates our correct
implementation and the feasibility of the imaginary-time method.

\begin{figure}[!t]
\centering
\includegraphics[width=0.85\linewidth,clip]{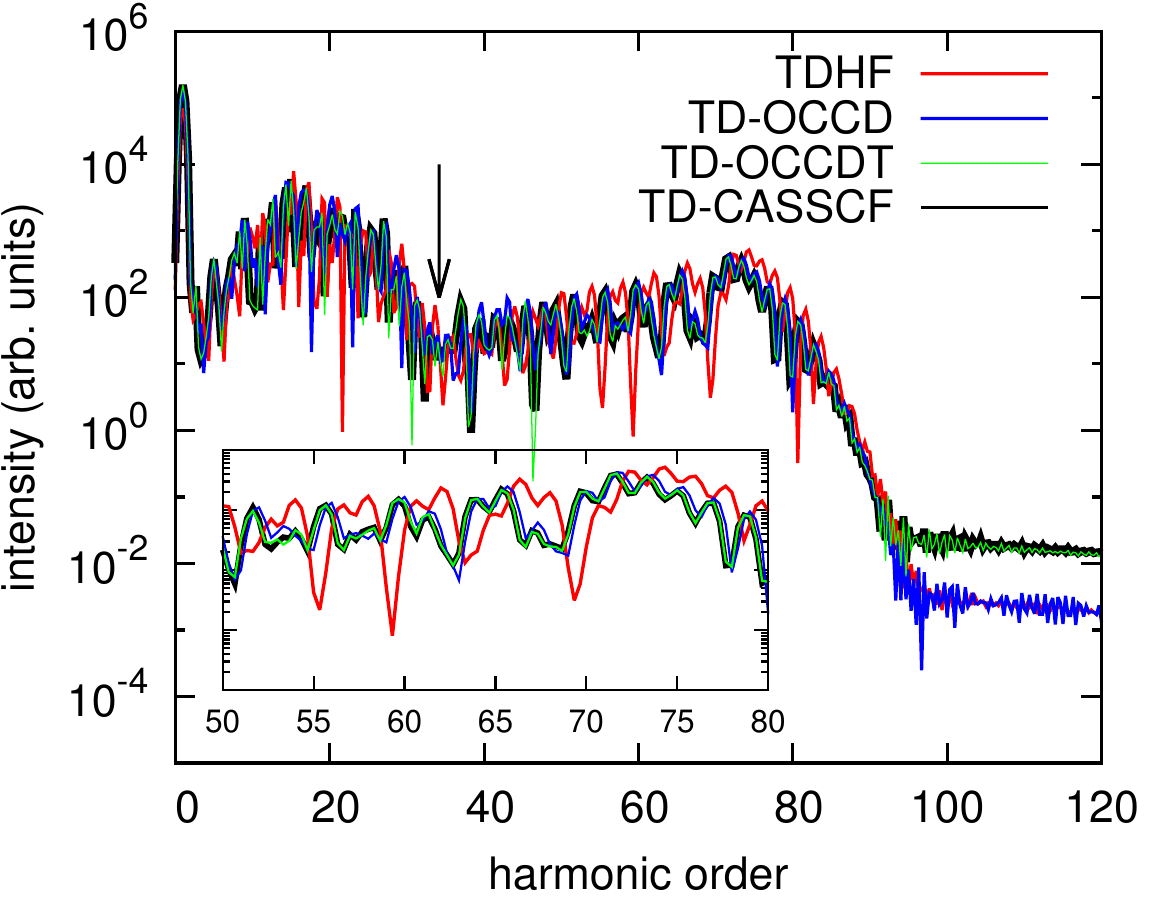}
\caption{\label{fig:hhg}
The HHG spectra of Ar exposed to a laser pulse with a wavelength of 800 nm and an intensity of
6$\times$10\ue{14} W/cm\ue{2}. Comparison of the results of TDHF, TD-OCCD,
TD-OCCDT, and TD-CASSCF methods. The inset shows a close-up of the spectra from 50th to 80th harmonic order.
}%
\end{figure}
\section*{Initial applications\label{sec:num}}

Finally we report our initial application of the TD-OCC method to 
Ar atom in a strong laser field.
We used the spherical FEDVR basis given as
the product $Y_{lm}(\theta,\phi)f_k(r)$ of spherical harmonics $Y_{lm}$ with $l\leq l_{\rm max}$ and 
the FEDVR basis $f_k(r)$, which divides the range of radial coordinate
$0<r<R_{\rm max}$ into finite elements of length 4 (with several
finer elements near the nuclei), each supporting 23 local DVR
functions. We first obtained the ground states of TDHF, TD-OCC,
and TD-CASSCF methods by the imaginary time relaxation. 
For the latter two methods, the Ne core was kept frozen at HF orbitals and
eight valence electrons were correlated among 13 optimized active orbitals, 
with their initial characters being $3s3p3d4p4s$.
\begin{figure}[!b]
\centering
\includegraphics[width=0.85\linewidth,clip]{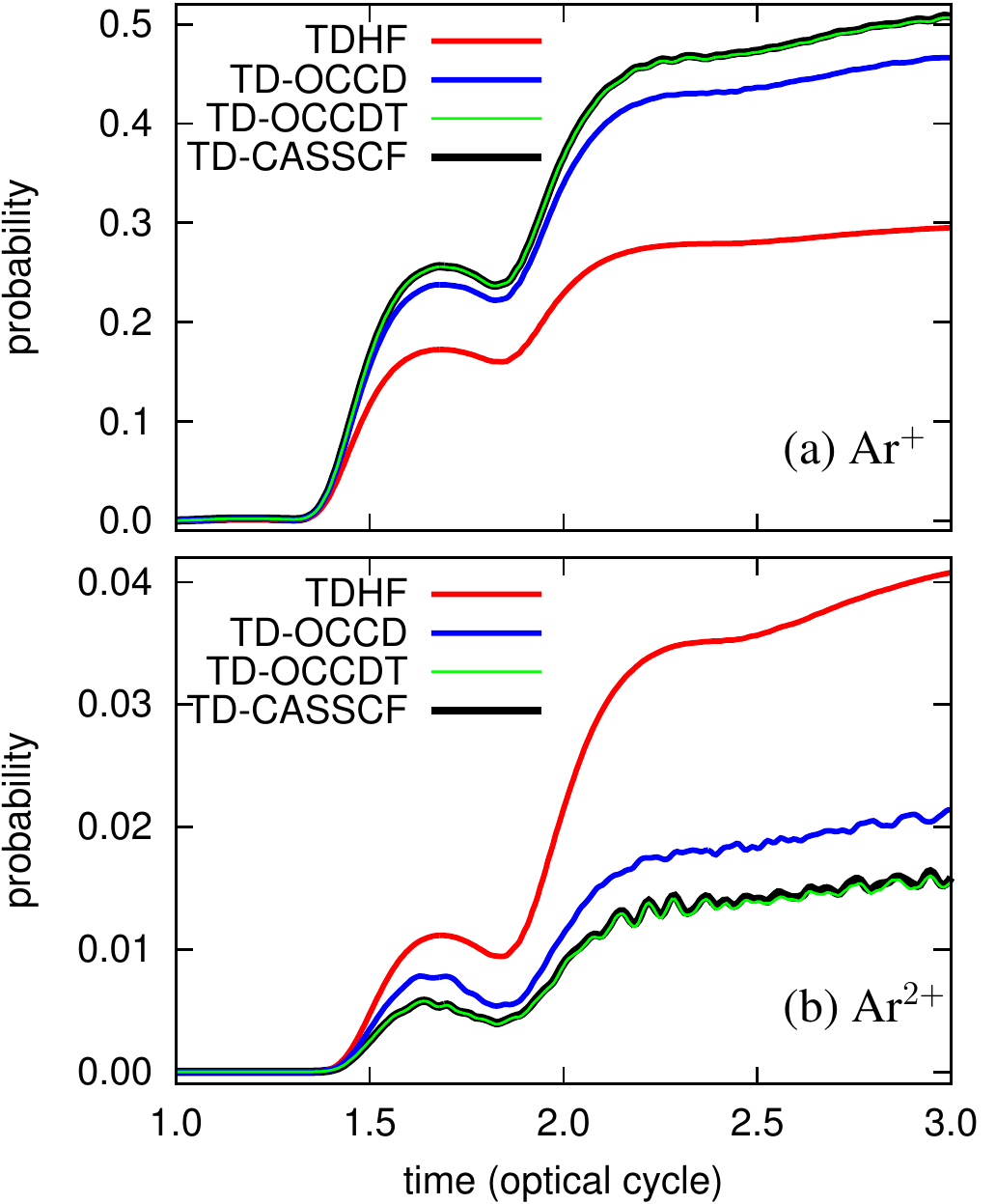}
\caption{\label{fig:ipd}
The probabilities, as a function of time, of finding one (a) and two (b) electrons outside a
sphere of radius $R_0 = 20$ a.u. Results of TDHF, TD-OCCD, TD-OCCDT, and
TD-CASSCF methods.
}%
\end{figure}
Starting from the ground states, we then simulated the electron
dynamics of Ar atom subject to a linearly polarized (in the $z$ direction) laser pulse
with a wavelength of 800 nm, a peak intensity of 6$\times$10\ue{14}
W/cm\ue{2}, and a foot-to-foot pulse duration of three optical cycles with a
sin\ue{2}-shaped envelope, within the dipole approximation in the velocity gauge.
The EOMs are propageted
using the fourth-order exponential Runge-Kutta method.\cite{Hochbruck:2010} The Ne core was frozen in all the methods
and 13 active orbitals were propagated in TD-OCC and TD-CASSCF methods. 
The basis parameters were calibrated to achieve
convergence with $R_{\rm max}=240$ (with the absorbing boundary\cite{Sato:2016}) and $l_{\rm max}=63$.

Figure~\ref{fig:hhg} shows high-harmonic generation (HHG) spectra calculated as the modulus squared of the 
Fourier transform of the expectation value of the dipole acceleration $\langle d^2\hat{z}/dt^2\rangle(t)$ which, in turn, 
is obtained with the modified Ehrenfest expression\cite{Sato:2016} using RDMs of Eq.~(\ref{eq:rdmh}).
All methods predicted qualitatively similar spectra, 
with a minimum at 53 eV ($\approx$ 34th order, indicated with an arrow) 
close to the Cooper minimum of photoionization spectrum,\cite{Cooper:1962} which was
experimentally observed.\cite{Worner:2009} The TDHF method, 
however, failed to reproduce fine structures of the TD-CASSCF spectrum especially at 
higher plateau as seen in the inset of Fig.~\ref{fig:hhg}. The description is largely improved by TD-OCCD, and the TD-OCCDT spectrum 
well reproduces the TD-CASSCF one in virtually all details.

We also compare the probabilities of finding one [Fig.~\ref{fig:ipd}~(a)] and two  [Fig.~\ref{fig:ipd}~(b)] electron(s) outside a sphere of radius $R_0=20$, 
which measure the single and double ionization probabilities, respectively.
These probabilites are much more sensitive to the description of correlated motions of electrons
than HHG,
which hinders a correct description with TDHF. 
The rapid converngence of TD-OCC results to the TD-CASSCF ones for such nonperturbatively 
nonlinear processes strongly promises the TD-OCC method to be 
a vital tool to investigate correlated high-field phenomena.

\section*{Summary\label{sec:sum}}
We have successfully formulated a new time-dependent coupled-cluster method called TD-OCC method, and 
implemented the first two, and the most important, series of approximations, TD-OCCD and TD-OCCDT. The present method is
size extensive and gauge invariant, a polynomial cost-scaling
alternative to the TD-MCSCF method. It would open new possibilities of high-accuracy first-principle investigations of
multielectron dynamics in ever-unreachable large target systems.
The rigorous derivation, details of the implementation, as well as 
other ansatz for TD-CC theories and comparison thereof, will be
presented in a forthcoming article.\cite{Sato:2018}

\begin{acknowledgments}
This research was supported in part by a Grant-in-Aid for Scientific Research 
(Grants No.~25286064, No.~26390076, No.~26600111, No.~16H03881, and 17K05070)
from the Ministry of Education, Culture, Sports, Science and Technology (MEXT) of Japan and also 
by the Photon Frontier Network Program of MEXT.
This research was also partially supported by the Center of Innovation Program from the Japan Science 
and Technology Agency, JST, and by CREST (Grant No.~JPMJCR15N1), JST.
Y.~O. gratefully acknowledges support from the Graduate School of
Engineering, The University of Tokyo, Doctoral Student Special
Incentives Program (SEUT Fellowship).
\end{acknowledgments}

\bibliography{refs.bib}
\end{document}